\documentclass[aps,twocolumn,
showpacs,pra,amssymb, amsmath]{revtex4}
\usepackage{amsmath,latexsym,amssymb}
\usepackage[dvips]{graphicx}
\usepackage{amsmath}
\usepackage{latexsym}
\usepackage{amssymb}
\usepackage{bm}

\newcommand{\one}{\mbox{$1 \hspace{-1.0mm}  {\bf l}$}}

\begin{document}

\title{Observables suitable for restricting the
fidelity to multipartite maximally entangled states}
\author{Koji Nagata, Masato Koashi, and Nobuyuki Imoto}
\affiliation{
CREST Research Team for Interacting Carrier Electronics,
School of Advanced Sciences, The Graduate University for
Advanced Studies (SOKEN), Hayama, Kanagawa, 240-0193, Japan
}
\date{\today}

\begin{abstract}

We present a class of observables which are suitable for 
determining the fidelity of a state to the multipartite
Greenberger-Horne-Zeilinger (GHZ) state. Given an 
expectation value of an observable belonging to the class,
we give a simple formula that gives a lower bound and 
an upper bound for the fidelity. 
Applying the formula to the GHZ-state
preparation  experiment by 
Pan {\it et al}. {[Nature (London) {\bf 403}, 515 (2000)]}, 
we show that the observed state lies outside of the
class of biseparable mixed three-qubit states. We also show 
that for this class of operators, 
adopting the principle of minimum variance 
{[Phys. Rev. A {\bf 60}, 4338 (1999)]} in the state estimation
always results in the state with the minimum fidelity.
\end{abstract}

\maketitle

\section{Introduction}

Recently, importance of entangled states
of multipartite system 
has been realized not only as a fundamental concept of quantum
mechanics
\cite{bib:GHZ} but also as an essential resource for quantum information
processing \cite{bib:QSS}.
 Up to now, there have already been several experimental 
reports on 3- and 4-particle entangled
states \cite{bib:dik,bib:pan,bib:GHZexp}.
One of the important measures 
to analyze how close the produced quantum state is 
to the desired maximally entangled 
state is the fidelity\cite{bib:fid}, i.e., the
overlap with the desired entangled state.
For bipartite systems and multipartite systems, the fidelity is used as 
a criterion for nonseparability
and distillability of the so-called Werner state or Werner-type state,
 i.e., a maximally entangled state mixed with the completely depolarized
state\cite{bib:sepadis}. 
In the classification of mixed three-qubit states\cite{bib:class}, 
it was shown that one of tripartite witnesses 
can be used to
detect a state that does not belong to the 
biseparable class. This
witness is given by 
\begin{eqnarray}
{\cal W}= \frac{1}{2}\one-P_{\rm GHZ}, \label{witness}
\end{eqnarray}
where $P_{\rm GHZ}$ is the projector onto a GHZ state.
When ${\rm Tr}[{\cal W}\ \rho]<0$, the state 
$\rho$ lies outside of the biseparable class and
has genuinely tripartite entanglement.
Since ${\rm Tr}[{\cal W}\ \rho]$ is written as
$1/2-f$ using the fidelity $f$ to the GHZ state, the 
fidelity is useful to determine to which 
class a state belongs\cite{bib:witope}.

As can be seen from these examples, it is important to
determine the fidelity $f$ of experimentally 
produced states from the observed data. 
Systematic arguments about the possible fidelity values
allowed by experimental data for multipartite system will 
be helpful to the experimental realization of 
various applications in quantum information processing.
The main purpose of this paper is to give a formula for 
the possible range of the fidelity value  
in the simplest case, i.e., the case where an expectation value of a 
single operator is given as experimental data.
We present a class of observables which are suitable for 
determining the range of the fidelity of a state 
to the $n$-partite GHZ state $|\Phi_n\rangle$.
 The class is determined 
through the expansion of the projection operator 
$|\Phi_n\rangle\langle \Phi_n|$ into
the sum of direct products of Pauli operators for 
each party.
Given an 
expectation value of an observable belonging to the class,
the formula gives a lower bound and 
an upper bound for the fidelity. 
As an example, we analyze
 the GHZ-state
preparation  experiment by Pan {\it et al}. \cite{bib:pan} 
and show that the fidelity to the GHZ state is larger than 
$0.71$. This indicates that
the observed state does not belong to the 
biseparable class and
has genuinely tripartite entanglement.


In addition to the argument of what kind of states are possible
under the constraints of experimental data, 
there also is a problem of determining which state is most likely
under the constraints. Such problems of estimation  
 for
bipartite system has been discussed 
 along the maximum entropy principle\cite{bib:Jaynes}.
Application of only this principle sometimes leads to an 
estimated state that possesses
stronger entanglement than the minimum entanglement that is compatible
with the measured data.
Based on the additional assumption that the realization
of a stronger entanglement is less
realistic,
Horodecki {\it et al}. introduced a new constraint, i.e., minimization of
entanglement\cite{bib:Horodecki} in applying the maximum entropy principle.
They thus obtained an estimated state that has the minimum entanglement.
Rajagopal derived the same state with a different assumption together with the
the maximum entropy principle, i.e., to minimize the variance of a Bell
operator\cite{bib:Rajagopal}.
Since then,
much attention has been paid\cite{bib:connect} to this problem.
Here we will show that, in multipartite systems, 
applying the minimum variance principle 
to the operators belonging to the above class gives the 
states with the minimum fidelity that is allowed by the 
constraints. This is a generalization of Rajagopal's 
results to multipartite systems, and reveals why and 
in what cases the minimum variance leads to small entanglement.


This paper is organized as follows. 
In Sec.~\ref{sec:desired},
we take a GHZ state as the state of interest
and present the class of operators by
decomposing the projector into the sum
of operators forming a commutative group.
In Sec.~\ref{sec:ineq}, we derive 
a simple formula that gives a lower bound and 
an upper bound for the fidelity. 
In Sec.~\ref{sec:exp}, we apply the formula to analyze
 the GHZ-state
preparation  experiment by Pan {\it et al}. \cite{bib:pan},
 and show that the fidelity to the GHZ state is larger than 
$0.71$.
Sec.~\ref{sec:vari} deals with the state estimation problem 
based on the minimum variance principle. 
Sec.~\ref{conclusion} concludes this paper.



\section{Decomposition of projector and class of
observables}\label{sec:desired}

In order to discuss the fidelity,
we have to specify one state of interest.
We have referred to the state as the {\it desired state}.
we take $n$-partite GHZ state $|\Phi_n\rangle$ as the desired state,
which is defined as 
\begin{eqnarray}
|\Phi_n\rangle:=\frac{1}{\sqrt{2}}(|+_1;+_2; \cdots ;+_n\rangle+
|-_1;-_2;\cdots;-_n\rangle).
\end{eqnarray}
In the following discussion, we consider the fidelity of the state 
$\rho$
to $|\Phi_n\rangle$,
i.e.,
$f:={\rm Tr}[\rho\ |\Phi_n\rangle\langle\Phi_n|]$.
The projection operator $|\Phi_n\rangle\langle\Phi_n|$ can be expanded as
\begin{eqnarray}
|\Phi_n\rangle\langle\Phi_n|&=&
\frac{1}{2^{n+1}}\Biggl(\prod^{n}_{j=1}(\sigma^j_x+i\sigma^j_y)+
\prod^{n}_{j=1}(\sigma^j_x-i\sigma^j_y)\nonumber\\
&&+\prod^{n}_{j=1}(I^j+\sigma^j_z)+
\prod^{n}_{j=1}(I^j-\sigma^j_z)\Biggl)\nonumber\\
&=&(1/2^{n})(O_0+O_1+\cdots+O_{2^{n}-1}),\label{NFO}
\end{eqnarray}
where $O_{p}$ is defined by
\begin{eqnarray}
O_{p}:=\prod_{j=1}^{n}(\sigma_x^j)^{b_0}(\sigma_z^j)^{b_j},\label{element}
\end{eqnarray}
where the $n$-bit sequence $b_0b_1\cdots b_{n-1}$
is the binary representation of
$p$, $b_n=\sum_{j=1}^{n-1}b_j$,  since 
the terms with odd parity for $b_1\cdots b_{n}$
vanish in the above expansion. 
The superscript $j$ of the Pauli operators  
denotes particle $j$.
It is easy
to see that $O_pO_q=O_{p\oplus q}$, where
$p\oplus q$ is the bitwise XOR of $p$ and $q$.
Hence the set of $2^n$ operators $\{O_p\}$ forms
a commutative group isomorphic to $(Z_2)^n$.
We denote this 
commutative group as 
$\Lambda_n$.
The operator $O_0$ is the identity operator for 
the $2^n$-dimensional space, and the other 
operators $O_1,\cdots,O_{2^{n}-1}$ 
have two eigenvalues, $\pm 1$.
All elements of
$\Lambda_n$ take $|\Phi_n\rangle$ as an eigenstate
with eigenvalue 1.

For $n=2$, the above expansion is explicitly written as follows,
\begin{eqnarray}
|\Phi_2\rangle\langle\Phi_2|&&=(1/4)
(I^1I^2+\sigma^1_z\sigma^2_z+i\sigma^1_y i\sigma^2_y+\sigma^1_x\sigma^2_x),
\label{2-fo}
\end{eqnarray}
where $I$ represents the identity operator for 
the $2$-dimensional space.
For $n=3$, it is written as
\begin{eqnarray}
&&|\Phi_3\rangle\langle \Phi_3|=(1/8)
({\cal O}_{III}+{\cal O}_{Izz}+{\cal O}_{zIz}+{\cal O}_{zzI}\nonumber\\
&&\quad
+{\cal O}_{xxx}+{\cal O}_{xyy}+{\cal O}_{yxy}+{\cal O}_{yyx}
),\label{3-fo}
\end{eqnarray}
where we have used the simplified notations as
${\cal O}_{III}:={ I}^1I^2I^3,\
{\cal O}_{Izz}:={ I}^1\sigma^2_z\sigma^3_z,\
{\cal O}_{zIz}:=\sigma^1_z{ I}^2\sigma^3_z,\
{\cal O}_{zzI}:=\sigma^1_z\sigma^2_z{ I}^3,\
{\cal O}_{xxx}:=\sigma^1_x\sigma^2_x\sigma^3_x,\
{\cal O}_{xyy}:=\sigma^1_x i\sigma^2_y i\sigma^3_y,\
{\cal O}_{yxy}:=i\sigma^1_y\sigma^2_x i\sigma^3_y$,
and
${\cal O}_{yyx}:=i\sigma^1_y i\sigma^2_y\sigma^3_x$.

Now let us consider the problem of determining the 
fidelity by measuring the expectation value of 
an observable. The most direct approach is, of course,
to measure $\langle|\Phi_n\rangle\langle\Phi_n|\rangle$,
which will be done by conducting $2^n-1$ different 
correlation measurements to determine 
$\langle O_j\rangle\ (j=1,2,\cdots,2^n-1)$. 
Our interest here is how we can deduce the information
about the fidelity from observables that can be measured
by much smaller number of correlation measurements.
In what follows, we give a formula to derive an inequality for 
the fidelity from an expectation value of an observable 
belonging to the class ${\cal C}_n$ defined as follows.

{\it Class} ${\cal C}_n$:  An observable ${\cal A}$ belongs to ${\cal C}_n$
if and only if ${\cal A}$ is a linear combination of 
operators $Q_1, Q_2, \cdots , Q_m\in \Lambda_n$ with
positive coefficients, and the set 
$\{Q_1, Q_2, \cdots , Q_m\}$ forms a
system of generators for $\Lambda_n$.

The class is determined through $\Lambda_n$, and hence 
determined by the desired state $|\Phi_n\rangle$.

The minimum cardinal number of systems of generators for $\Lambda_n$ 
is $n$.
To see this,
suppose there are only
$l(<n)$ generators.
In this case, however,
they can generate at most 
$\sum^l_{i=0} {_l}C_i=2^{l}(<2^{n})$
kinds of elements of $\Lambda_n$ due to the property of the Pauli matrices.
An example of a system of generators for $\Lambda_n$ is
\begin{eqnarray}
\{O_{x,x,\cdots,x},\overbrace{O_{z,I,I,\cdots,z},
O_{I,z,I,I,\cdots,z},\cdots,O_{I,I,\cdots,z,z}}^{n-1}\},
\end{eqnarray}
where
$O_{x,x,\cdots,x}:=\sigma^1_x\sigma^2_x\cdots\sigma^n_x$,
$O_{I,z,I,\cdots,z}:={ I}^1\sigma^2_z{ I}^3\cdots\sigma^n_z$,
and so on\cite{bib:Buzek}.
This system of generators for $\Lambda_n$, 
indeed, generates all $2^n$ elements of 
$\Lambda_n$ with the help of Eq.~(\ref{element}).

As for $\Lambda_2$,
examples of a system of generators are
$\{\sigma^1_x\sigma^2_x,\ \sigma^1_z\sigma^2_z\}$, 
$\{\sigma^1_x\sigma^2_x,\ \sigma^1_z\sigma^2_z,\ i\sigma^1_y i\sigma^2_y\}$, 
$\{I,\ \sigma^1_x\sigma^2_x,\ i\sigma^1_y i\sigma^2_y\}$ 
and so on.
As for $\Lambda_3$, 
examples of a system of generators are
$\{{\cal O}_{xyy}$, ${\cal O}_{yxy}$, ${\cal O}_{yyx}\}$, 
$\{{\cal O}_{xyy}$, ${\cal O}_{yxy}$, ${\cal O}_{Izz}\}$, 
$\{{\cal O}_{xxx}$, ${\cal O}_{xyy}$, ${\cal O}_{yxy}$, ${\cal O}_{yyx}\}$ 
and so on.
\section{Inequality for fidelity}\label{sec:ineq}

In this section, we derive an inequality under the 
condition that the expectation value of an operator 
${\cal A}\in {\cal C}_n$ is specified. First, we will show that
if the given expectation value is the maximum value, 
the state must be $|\Phi_n\rangle$. For that, we use 
the following lemma.

{\it Lemma.} Let $X,\ Y$ be operators taking
 eigenvalues $\pm 1$,
and $[X,Y]=0$.
Then,
\begin{eqnarray}
1-|\langle X\rangle-\langle Y\rangle|
\geq\langle XY\rangle\geq\langle X\rangle+\langle Y\rangle-1.\label{cov}
\end{eqnarray}
This is directly proven by inequalities:
\begin{eqnarray}
\langle(1-X)(1-Y)\rangle\geq 0 \quad {\rm and} \quad
\langle(1\pm X)(1\mp Y)\rangle\geq 0.
\end{eqnarray}

In the following,
we consider 
${\cal A}=\alpha_1 Q_1+\alpha_2 Q_2+\cdots+\alpha_m Q_m$ 
$(2^{n}\geq m \geq n,\ \alpha_i>0,\ \forall\ i)$.
We assume that 
${\cal A}\in {\cal C}_n$.
We assume that the maximum expectation value
$\langle{\cal A}\rangle=M$ is given,
where $M:=\sum^m_{i=1} \alpha_i$ is the maximum eigenvalue of ${\cal A}$.
Because all coefficients $\alpha_i$ are positive, $\langle{\cal A}\rangle=M$ 
implies $\langle Q_1\rangle=\langle Q_2\rangle=\cdots=\langle Q_m\rangle=1$.
Since $Q_j^2=I$ for any $j$ and the set 
$\{Q_1, Q_2, \cdots , Q_m\}$ forms a
system of generators for $\Lambda_n$, any element $Q$ of $\Lambda_n$ 
can be written as
\begin{eqnarray}
Q=Q_{\beta_1}Q_{\beta_2} \cdots Q_{\beta_l},
\end{eqnarray}
with $1\le\beta_1 < \beta_2 <\cdots <\beta_l \le m$.
We show that $\langle Q\rangle=1$ for all $l$ as follows.
When $l=1$, then $\langle Q\rangle=1$ holds.
Suppose $\langle Q\rangle=1$ holds when $l=k$, i.e., 
$\langle\prod^k_{i=1} Q_{\beta_i}\rangle=1$.
Then with the help of the lemma, 
$\langle\prod^{k+1}_{i=1} Q_{\beta_i}\rangle=1$ holds.
Let $X$ be $\prod^k_{i=1} Q_{\beta_i}$ and $Y$ be $Q_{\beta_{k+1}}$, 
respectively.
The lemma leads that 
$1\leq \langle \prod^{k+1}_{i=1} Q_{\beta_i}\rangle\leq 1$.
Hence $\langle Q_1\rangle=\langle Q_2\rangle=\cdots=\langle Q_m\rangle=1$ means the expectation values of 
all elements of 
$\Lambda_n$ are one.
This means 
that the fidelity is 1 with the help of Eq.~(\ref{NFO}).
Therefore we obtain the following: 

{\it Proposition 1}: Let $M$ be the largest eigenvalue of 
${\cal A}\in {\cal C}_n$. If ${\rm Tr}[\rho\ {\cal A}]=M$,
then $\rho=|\Phi_n\rangle\langle\Phi_n|$.

This implies that
the largest eigenvalue $M$ is not degenerate.
This is crucial point in deriving an
inequality for the fidelity.
Let us write the eigenvalues of
${\cal A}$ as
$M,\ r_2,\ r_3,\cdots,$ and
$r_s$($s\leq 2^n$), where
$M>r_2>r_3>\cdots>r_s\geq -M$.
In this notation,
when some eigenvalues of
${\cal A}$ are degenerate, then
$s< 2^n$ holds, and when all the eigenvalues of
${\cal A}$ are not degenerate, then
$s=2^n$ holds.
Since proposition 1 implies that $|\Phi_n\rangle$ is the 
only eigenstate for the largest eigenvalue $M$, 
we can generally expand $\langle {\cal A}\rangle:={\rm Tr}[\rho\ {\cal
A}]$ as 
\begin{equation}
\langle {\cal A}\rangle=Mf+\sum^s_{i=2} q_i r_i,
\end{equation}
where $f={\rm Tr}[\rho\ |\Phi_n\rangle\langle\Phi_n|]$ is the 
fidelity to $|\Phi_n\rangle$, and $q_i(\ge 0)$ satisfy
$f+\sum^s_{i=2} q_i=1$. Using this relation to eliminate 
$q_2$, we have 
\begin{eqnarray}
\langle{\cal A}\rangle&=&Mf+(1-f-\sum^s_{i=3} q_i)r_2+
\sum^s_{i=3} q_i r_i
\nonumber\\
&=&(M-r_2)f+r_2-
\sum^s_{i=3} q_i (r_2-r_i),
\end{eqnarray}
We thus obtain
\begin{equation}
\frac{\langle{\cal A}\rangle-r_2}{M-r_2}\leq f.
\label{n-ineq}
\end{equation}
The equality of the relation (\ref{n-ineq}) holds when
$\sum^s_{i=3} q_i=0$.

We can also derive an inequality that gives an
upper bound of the fidelity by eliminating $q_s$, namely,
\begin{eqnarray}
\langle{\cal A}\rangle&=&Mf+\sum^{s-1}_{i=2} q_i r_i+
(1-f-\sum^{s-1}_{i=2} q_i)r_s
\nonumber\\
&=&(M-r_s)f+
\sum^{s-1}_{i=2} q_i (r_i-r_s)+r_s,
\end{eqnarray}
and
\begin{equation}
\frac{\langle{\cal A}\rangle-r_s}{M-r_s}
\geq f.
\label{n-inequp}
\end{equation}
The equality of the relation (\ref{n-inequp}) holds when
$\sum^{s-1}_{i=2} q_i=0$. 
We therefore obtain the following proposition.

{\it Proposition 2}: Let $M$, $r_2$, and $r_s$ be 
the largest, the second-largest, and the smallest eigenvalue
of ${\cal A}\in {\cal C}_n$, respectively. 
When $\langle{\cal A}\rangle:={\rm Tr}[\rho\ {\cal A}]$
is given, the fidelity  
$f:={\rm Tr}[\rho\ |\Phi_n\rangle\langle\Phi_n|]$ is bounded as 
\begin{eqnarray}
\frac{\langle{\cal A}\rangle-r_2}{M-r_2}\leq f 
\leq \frac{\langle{\cal A}\rangle-r_s}{M-r_s}\ .\label{final}
\end{eqnarray}

\section{application to experimental data}\label{sec:exp}

We analyze the experimental data 
by Pan {\it et al}.\cite{bib:pan}.
In this experiment they obtained 
four expectation values of three-photon polarization 
correlations,
\begin{eqnarray}
&&\langle xyy\rangle\simeq
\langle yxy\rangle\simeq
\langle yyx\rangle\simeq 0.70, \nonumber\\
&&\langle xxx\rangle\simeq 0.74. \label{experidata0}
\end{eqnarray}
These experimental data is obtained by the post selection, i.e., 
picking up only the events with each of the three detectors 
registering a photocount. If the detectors used in the experiments 
were ideal ones, the post-selected state would be 
contained in a $2^3$-dimensional subspace, which can be
identified with a tripartite system of three qubits. 
Then the above observed values could be considered to give
the expectation values 
$\langle{\cal O}_{xyy}\rangle$,
$\langle{\cal O}_{yxy}\rangle$,
$\langle{\cal O}_{yyx}\rangle$,
and $\langle{\cal O}_{xxx}\rangle$.
 In the real experiment, however, the detectors are not ideal, 
namely, they cannot distinguish a single photon 
from more than one photons and they have a limited quantum
efficiency and dark counting. 
Due to these imperfection together with the nonideal photon source,
the post-selected state also contains contributions
outside of the $2^3$-dimensional subspace, in
which two photons or no photons enter the same detector. 
However, the superfluous contributions can be neglected 
as compared to the statistical uncertainty (a few \%) of the observed 
values as follows\cite{bib:dik}.
The contribution of no-photon events are due to 
dark counting, but 
the rate of the dark counts is low enough \cite{bib:dik} to be able 
to neglect the effect.
The contribution of more than one photon entering a detector
passes the post selection only if another detector have a 
dark count, or more than two photon pairs are created in the
parametric downconversion. The former case is negligible 
due to the low dark count rate, and the latter is also 
negligible since the probability per pulse to create 
$n$-photon pairs is of the order of about $10^{-4n}$.
We can thus 
assume that the post-selected state is related to polarization of three
photons and approximately supports $2^3$-dimensional Hilbert space. Hence
we obtain 
\begin{eqnarray}
&&\langle{\cal O}_{xyy}\rangle\simeq
\langle{\cal O}_{yxy}\rangle\simeq
\langle{\cal O}_{yyx}\rangle\simeq 0.70, \nonumber\\
&&\langle{\cal O}_{xxx}\rangle\simeq 0.74. \label{experidata}
\end{eqnarray}
The limited quality of the polarization optics just before the 
detectors may make the visibility lower, which will 
make the estimated fidelity smaller.
Hence we use these experimental expectation values 
for restricting the fidelity from below. 

Clearly, each observable of 
(\ref{experidata}) does not belong to ${\cal C}_3$.
Therefore we take the summation of these 
expectation values. 
We then obtain
\begin{eqnarray}
\langle{\cal A}\rangle
\simeq 2.84, \label{experidata2}
\end{eqnarray}
where ${\cal A}=
{\cal O}_{xyy}+{\cal O}_{yxy}+{\cal O}_{yyx}+{\cal O}_{xxx}$.
Apparently, 
${\cal A}\in{\cal C}_3$ holds, and eigenvalues of ${\cal A}$
are $4,\ 0$, and $-4$, where $0$ is a degenerate eigenvalue.
With the help of Eq.~(\ref{final}), 
where the parameters are set to 
$M=4$ and $r_2=0$, 
we can state that the observed state in this experiment 
has the fidelity to a GHZ state 
larger than or equal to $0.71$\cite{bib:sufficient}.
Because the value is lager than 1/2, with the help of Eq.~(\ref{witness}), 
these experimental data, indeed, ensure 
the observed state does not belong to the biseparable class and
has genuinely tripartite entanglement.

\section{relation between variance and fidelity}\label{sec:vari}

In this section, 
we show that 
if we apply the minimum variance principle 
for estimating the states from an experimentally obtained 
expectation value of an operator belonging to ${\cal C}_n$,
the estimated fidelity becomes the minimum value that is allowed 
by the constraints. This is a generalization of Rajagopal's 
results \cite{bib:Rajagopal} to multipartite systems.

Let us consider Rajagopal's case, i.e., bipartite system 
in which the variance of Bell-CHSH operator 
$B=\sqrt{2}(\sigma^1_x\sigma^2_x+\sigma^1_z\sigma^2_z)$ 
is made minimal.

Suppose that an expectation value
$\langle{\cal A}\rangle$
is given, where
\begin{eqnarray}
&&{\rm Case\ 1}\quad{\cal A}=\sigma^1_x\sigma^2_x,\nonumber\\
&&{\rm Case\ 2}\quad{\cal A}=\sigma^1_x\sigma^2_x+\sigma^1_z\sigma^2_z.
\end{eqnarray}
Note that, for Case 1, the operator ${\cal A}$ does not belong to 
${\cal C}_2$, 
whereas, for Case 2, the operator ${\cal A}$ belongs to ${\cal C}_2$,
and ${\cal A}=B/\sqrt{2}$.

For Case 2,
combining the results by Refs.\cite{bib:Horodecki} and \cite{bib:Rajagopal},
it is shown that if variance
$\langle(\Delta{\cal A})^2\rangle$ is made minimal,
the calculated fidelity to $|\Phi_2\rangle$ takes minimal
when the expectation value $\langle{\cal A}\rangle$ ($\geq 0$) is given.
In this way,
the minimum entangled state was derived from Jaynes principle.
For Case 1, the given
$\langle{\cal A}\rangle$ determines a range
$0 \le f \le \frac{\langle {\cal A}\rangle +1}{2}$
for the possible value of $f$,
and the minimization condition for
$\langle(\Delta{\cal A})^2\rangle$
puts no further condition on this range.
The allowed fidelity value is thus unsettled and can be any value in the
region, except for the case that 
$\langle{\cal A}\rangle= -1\ (\rightarrow f=0)$. 
(Remember $-1 \leq\langle{\cal A}\rangle\leq 1$ for Case 1). 
This also means that, for Case 1, we cannot derive minimum fidelity (i.e., 
zero) from the minimum variance principle.

Next we consider several examples for tripartite system.
There are eight GHZ states,
which are written as
\begin{eqnarray}
&&|\psi_{1(8)}\rangle:=(1/\sqrt{2})
    (|+_1;+_2;+_3\rangle\pm|-_1;-_2;-_3\rangle),\nonumber\\
&&|\psi_{2(7)}\rangle:=(1/\sqrt{2})
    (|-_1;+_2;+_3\rangle\pm|+_1;-_2;-_3\rangle),\nonumber\\
&&|\psi_{3(6)}\rangle:=(1/\sqrt{2})
    (|+_1;-_2;+_3\rangle\pm|-_1;+_2;-_3\rangle),\nonumber\\
&&|\psi_{4(5)}\rangle:=(1/\sqrt{2})
    (|+_1;+_2;-_3\rangle\pm|-_1;-_2;+_3\rangle).
\end{eqnarray}
The eight probabilities of observation of these GHZ states 
for the state $\rho$ are defined as
\begin{eqnarray}
p_{i}:= \langle\psi_{i}|\ \rho\ |\psi_{i}\rangle,\quad(i=1,2,\cdots,8).
\end{eqnarray}
We can see that $|\psi_1\rangle$ is equal to $|\Phi_3\rangle$.
The fidelity $f$ is then identical to $p_1$.

Suppose that an expectation value
$\langle{\cal A}\rangle$
is given,
where we consider
\begin{eqnarray}
&&{\rm Case\ 1}\quad{\cal A}={\cal O}_{xyy}+{\cal O}_{yxy},
\nonumber\\
&&{\rm Case\ 2}\quad{\cal A}={\cal O}_{xyy}+{\cal
O}_{yxy} +{\cal O}_{zzI},\nonumber\\
&&{\rm Case\ 3}\quad{\cal A}={\cal O}_{xyy}+{\cal
O}_{yxy} +{\cal O}_{yyx}.
\end{eqnarray}
Note that, for Case 1 and 2, 
the operator ${\cal A}$ does not belong to ${\cal C}_3$, 
whereas, for Case 3, the operator ${\cal A}$ belongs to ${\cal C}_3$.

For Case 3, 
when the variance $\langle(\Delta {\cal A})^2\rangle$
is made minimal,
we see, later, that we can derive the minimum fidelity.
For Case 1 and 2, however,
the minimum variance principle does not work completely.
In Case 1 and 2, the given
$\langle{\cal A}\rangle$ determines range 
$0 \leq f \leq \frac{\langle {\cal A}\rangle+2}{4}$ and 
$0 \leq f \leq \frac{\langle {\cal A}\rangle+1}{4}$, respectively
for the possible value of $f$.
For Case 2, the minimization condition for
$\langle(\Delta{\cal A})^2\rangle$
puts no further condition on this range\cite{bib:co.}.
For Case 1,
the minimization of
$\langle(\Delta{\cal A})^2\rangle$ makes
$f_{\rm max}$ smaller, i.e., the allowed fidelity value 
is 0 for $-2\leq \langle{\cal A}\rangle\leq 0$ 
and $0 \le f \le \frac{\langle {\cal A}\rangle}{2}$ 
for $0< \langle{\cal A}\rangle\leq 2$. 
(Remember $-2\leq \langle{\cal A}\rangle\leq 2$ for Case 1).
This means that, for Case 2, we cannot 
derive the minimum fidelity (i.e., 
zero) from the minimum variance principle 
and for Case 1, we cannot 
derive the minimum fidelity for 
$0< \langle{\cal A}\rangle\leq 2$.

Now we calculate the fidelity for Case 3 from the minimum 
variance principle.
If we write operator
${\cal A}$ in the matrix form using the GHZ basis,
the diagonal elements become
$3,\ -1,\ -1,\ -1,\ 1,\ 1,\ 1,$ and $-3$,
and no off-diagonal element appears.
This means that the measured value for
${\cal A}$ can take four values
$3,\ -1,\ 1,$ and
$-3$,
where $-1$ and 1 are degenerate eigenvalues.
Using notations
$p_{\alpha}:=p_2+p_3+p_4 ,\ p_{\beta}:=p_5+p_6+p_7$,
and the relation
$f=p_1$,
the probability that the measured value for
${\cal A}$ takes 3,\  1,\  $-1$, or $-3$ is expressed as
\{$f, p_{\beta}, p_{\alpha}, p_8$\}.
We can calculate this for the three cases corresponding to
$3 \ge \langle{\cal A}\rangle > 1$,
$1 \ge \langle{\cal A}\rangle > -1$, and
$-1 \ge \langle{\cal A}\rangle \ge -3$,
as follows.
When $\langle{\cal A}\rangle$ lies between 3 and 1, the measured value for
${\cal A}$ can take only 3 or 1 but not $-1$ or $-3$ to attain the
minimum variance of its distribution.
The minimization of
$\langle(\Delta{\cal A})^2\rangle$
thus leads to the distribution
\{$f,\ p_{\beta},\ p_{\alpha},\ p_8$\}
to be
$\left\{ \frac{\langle {\cal A}\rangle-1}{2}, \frac{3-\langle
{\cal A}\rangle}{2}, 0, 0\right\}$. Similarly, for
$1 \ge \langle{\cal A}\rangle > -1$ case,
the distribution is calculated to be
$\left\{ 0,\frac{1+\langle {\cal A}\rangle}{2}, \frac{1-\langle {\cal 
A}\rangle}{2},
0\right\}$, and for
$-1 \ge \langle{\cal A}\rangle \ge -3$ case to be
$\left\{ 0, 0, \frac{3+\langle {\cal A}\rangle}{2}, \frac{-1-\langle
{\cal A}\rangle}{2}\right\}$.  The derived fidelity in Case 3,
by the minimum variance principle,
is then summarized as
\begin{eqnarray}
f=\left\{
\begin{array}{cl}
\displaystyle
\frac{\langle {\cal A}\rangle-1}{2}
& \quad 3\geq \langle{\cal A}\rangle>1\\
\\
\displaystyle
0
& \quad 1\geq \langle{\cal A}\rangle\geq -3 .
\end{array} \right.\label{fid-3}
\end{eqnarray}
It is easy to show that this is equal to the minimum of the fidelity values 
with the help of Eq.~(\ref{final}), 
where the parameters are set to be
$M=3$ and $r_2=1$.
It is thus concluded that, for Case 3, the derived fidelity to
$|\Phi_3\rangle$
by the minimum variance principle leads to the minimum 
of the possible fidelity values allowed by the expectation value.



We generalize the argument for 
$n$-partite system.
We consider 
${\cal A}=\alpha_1 Q_1+\alpha_2 Q_2+\cdots+\alpha_m Q_m\in {\cal C}_n$.
Suppose that an expectation value
$\langle{\cal A}\rangle$ is given.
In the following we calculate the fidelity $f$ from 
the minimum variance principle in this case.
We write the eigenvalues of
${\cal A}$ as
$M,\ r_2,\ r_3,\cdots,$ and
$r_s$, where
$M>r_2>r_3>\cdots>r_s\geq -M$. 
The probabilities for observing these eigenvalues,
$M,\ r_2,\ r_3,\cdots,$ and $r_s$ are denoted as
$f,\ q_2,\ q_3,\cdots,$ and $q_s$, respectively.
Similarly to the discussion as to Case 3 for tripartite system,
if $\langle{\cal A}\rangle$ lies between
$M$ and
$r_2$, the minimization of
$\langle(\Delta {\cal A})^2\rangle$ leads to
$\sum^s_{i=3} q_i = 0$, which means
$\{f,\ q_2,\ q_3,\cdots, q_s\}$ =
$\left\{ \frac{\langle {\cal A}\rangle-r_2}{M-r_2}, \frac{M-\langle
{\cal A}\rangle}{M-r_2}, 0, \cdots, 0 \right\}$.
The derived fidelity is then summarized as
\begin{eqnarray}
f=\left\{
\begin{array}{cl}
\displaystyle
\frac{\langle {\cal A}\rangle-r_2}{M-r_2}
&\quad M\geq \langle{\cal A}\rangle>r_2\\
\\
\displaystyle
0
&\quad r_2\geq \langle{\cal A}\rangle\geq r_s.
\end{array} \right.\label{fid-n}
\end{eqnarray}
Eq.~(\ref{fid-3}) is a special case of Eq.~(\ref{fid-n}) where the parameters are set to be
$M=3,\ r_2=1$ and $r_s=-3$.
We can see that Eq.~(\ref{fid-n}) gives the minimum fidelity 
with the help of Eq.~(\ref{final})\cite{bib:max}.
Hence we have the following result: 

{\it Proposition 3}: When $\langle{\cal A}\rangle:={\rm Tr}[\rho\ {\cal A}]$
is given, where ${\cal A}\in {\cal C}_n$, 
the derived fidelity to $|\Phi_n\rangle$ 
from the minimum variance principle is the minimum
of the possible fidelity values allowed by the expectation value.




\section{conclusion}\label{conclusion}

In conclusion,
we have analyzed the possible fidelity values 
that are compatible with an expectation value of a 
single operator as experimental data.
We have defined the desired maximally entangled state and 
formulated one class that 
is related to a decomposition of the projector onto the 
desired state.
We have made use of
the commutative group theory to formulate a 
class of observables. 
When an expectation value of an operator 
that belongs to the class is given, 
we can derive an inequality that gives
a lower bound and an upper bound of the fidelity 
values that are compatible with the expectation value.
With the help of the inequality, 
we have analyzed the experimental data 
by Pan {\it et al}.\cite{bib:pan}. 
The data ensure the observed state does not belong to 
the biseparable class and
has genuinely tripartite entanglement.
Finally, we have also analyzed the calculated 
fidelity from the minimum variance principle.

\acknowledgements

The authors are very grateful to 
S. Takagi for calling our attention to the present problem, 
and also 
A. Miranowicz for 
his intensive reading of the manuscript.
This work was partly supported by the Grant-in-Aid for Scientific
Research (B) (Grant No. 12440111) by Japan Society of the Promotion of
Science.


\begin{thebibliography}{9}
\bibitem{bib:witope}
The states beyond the biseparable class
are further classified into
the GHZ class and the W class.
The GHZ witness
is given by 
\begin{eqnarray}
{\cal W}_{\rm GHZ}= \frac{3}{4}\one-P_{\rm GHZ}.
\end{eqnarray}
When ${\rm Tr}[{\cal W}_{\rm GHZ}\ \rho]<0$, the state 
$\rho$ belongs to the GHZ class and not to the W class.
See \cite{bib:class}.
\bibitem{bib:co.}
This is due to the following fact: 
The measured value for 
${\cal A}={\cal O}_{xyy}+{\cal
O}_{yxy} +{\cal O}_{zzI}$ can take only two values
$3$ and $-1$,
where both of them are degenerate eigenvalues.
\bibitem{bib:max}
Incidentally, the maximization of $\langle(\Delta {\cal A})^2\rangle$ 
leads the distribution $\{f,\ q_2,\ q_3,\cdots, q_s\}$, to be 
$\left\{ \frac{\langle {\cal A}\rangle-r_s}{M-r_s}, 0, 
\cdots, 0, \frac{M-\langle{\cal A}\rangle}{M-r_s} \right\}$.
We can see that this gives the maximum fidelity 
with the help of Eq.~(\ref{final}).
\end{thebibliography}
\end{document}